\documentclass[12pt]{emulateapj}
\usepackage{blindtext}
\usepackage{float}
\usepackage{placeins}
\usepackage{graphicx}
\usepackage{amssymb, amsmath}
\usepackage{natbib}
\usepackage[caption=false]{subfig}
\DeclareMathOperator{\sech}{sech}
\begin{document}
\title{The Properties of Reconnection Current Sheets in GRMHD Simulations of Radiatively Inefficient Accretion Flows}

\author{David Ball\altaffilmark{1}, Feryal \"Ozel\altaffilmark{1}$^{,}$\altaffilmark{2}, Dimitrios Psaltis\altaffilmark{1}$^{,}$\altaffilmark{3}, Chi-Kwan Chan\altaffilmark{1}, Lorenzo Sironi\altaffilmark{4}}

\altaffiltext{1}{Department of Astronomy and Steward Observatory, University of Arizona, 933 N. Cherry Avenue, Tucson, Arizona 85721, USA}
\altaffiltext{2}{Guggenheim Fellow}
\altaffiltext{3}{Radcliffe Institute for Advanced Study, Harvard University, Cambridge, MA 02138, USA}
\altaffiltext{4}{Department of Astronomy, Columbia University, 550 West 120th Street, New York, New York 10027, USA}

\begin{abstract}
Non-ideal MHD effects may play a significant role in determining the dynamics, thermal properties, and observational
signatures of radiatively inefficient accretion flows onto black holes. In particular, particle acceleration during magnetic
reconnection events may influence black hole spectra and flaring properties. We use representative GRMHD simulations
of black hole accretion flows to identify and explore the structures and properties of current sheets as potential sites of magnetic
reconnection. In the case of standard and normal (SANE) disks, we find that, in the reconnection sites, the plasma beta ranges
from $0.1$ to 1000, the magnetization ranges from $10^{-4}$ to 1, and the guide fields are weak compared to the reconnecting
fields. In magnetically arrested (MAD) disks, we find typical values for plasma beta from $10^{-2}$ to $10^3$, magnetizations
from $10^{-3}$ to 10, and typically stronger guide fields, with strengths comparable to or greater than the reconnecting fields.
These are critical parameters that govern the electron energy distribution resulting from magnetic reconnection and can be
used in the context of plasma simulations to provide microphysics inputs to global simulations. We also find that ample magnetic
energy is available in the reconnection regions to power the fluence of bright X-ray flares observed from the black hole in the
center of the Milky Way.

\end{abstract}
\keywords{Accretion, magnetic reconnection --- acceleration of particles --- X-ray flares --- supermassive black holes --- Sgr A*} 
\maketitle

\section{Introduction}
General relativistic magnetohydrodynamic (GRMHD) simulations are often
used to study the physics of accretion systems around compact objects
and to explain their observed properties.  
These simulations make a number of simplifications that reduce
computation time and facilitate calculations.  The most common is the assumption of
``ideal'' MHD, which enforces that the plasma is infinitely conductive.  This
assumption leads to a couple of important properties: any electric
fields in the fluid frame are shorted out and, as a result, the
magnetic fields are frozen into the fluid (see, e.g.,
\citealt{Kulsrud2005}).  

For many systems, the ideal MHD approximation is adequate:
astrophysical plasmas are often fully ionized and have extremely low
resistivities.  However, even in systems where the approximation may
globally seem appropriate, there can arise regions that violate
the underlying assumptions.  One such example is when two fluid elements with
opposing magnetic fluxes encounter each-other.  Conventional wisdom tells
us that this may be a site of reconnection, where magnetic fields can
change topology and dissipate their energy into the plasma.  Such a
configuration will have a rapidly changing magnetic field in space,
resulting in a high current density.  Ohmic dissipation, which scales as $\eta
J^{2}$, where $\eta$ is the resistivity and J is the current density, can lead to significant dissipation where the current is high enough.  This shows that dissipative terms can change the energetics of the flow and, given their localization, can lead to time-dependent phenomena that are not captured in global ideal MHD simulations.

In the case of magnetic reconnection, even the inclusion of non-ideal
terms may not be sufficient to capture the entire behavior of the
plasma.  For instance, reconnection has been shown to be an efficient source of non-thermal particle acceleration under certain conditions
(\citealt{drake2013}; \citealt{sironi2014}; \citealt{melzani2014}; \citealt{liguo2015}; \citealt{guo2015}; \citealt{werner2016}; \citealt{sironi2016}; see \citealt{kagan2015} for review), but the common use of the fluid
equations assumes that the particle distribution is a
Maxwellian.  As a result, typical implementations of MHD do not capture effects related to non-thermal
particles, regardless of the inclusion of non-ideal terms.  In these
cases, we must turn to computational methods of solving the Vlasov
equation that do not make assumptions about the particle distribution.

The supermassive black hole at the center of the Milky Way, Sgr~A*,
has an accretion flow that falls into a category broadly referred to
as Radiatively Inefficient Accretion Flows (see \citealt{yuan2014}
for a recent review).  These flows are characterized by
geometrically thick, optically thin disks, low accretion rates, and
low luminosities.  In recent years, there have been a number of studies using GRMHD to
infer properties of the accretion flow around Sgr~A* (e.g., \citealt{narayan2012}; \citealt{dexter2012}; \citealt{drappeau2013};
\citealt{chan2015a}; \citealt{moscibrodzka2014}).  Even more recently, studies have begun to evolve the electron entropy equation, accounting for electron heating and anisotropic conduction (\citealt{ressler2015},
2016).  Additionally, \citet{chael2017} developed a scheme for coevolving a population of non-thermal electrons, including effects of adiabatic compression and expansion as well as radiative cooling.  While these simulations successfully match a
number of broadband steady-state properties, they show very little
X-ray variability, contrary to observations (\citealt{eckart2004};
\citealt{eckart2006}; \citealt{neilsen2013}, 2015).
In \citet{ball2016}, we showed that a population of non-thermal
electrons in highly magnetized regions of a radiatively inefficient accretion flow studied with GRMHD, where they are likely
to be accelerated via reconnection, can result in X-ray variability
with properties that are roughly consistent with observations. 

In this paper, we use representative GRMHD simulations to assess whether reconnection regions frequently occur in global simulations.  We consider simulations with Standard and Normal Evolution (SANE) and Magnetically Arrested Disk (MAD) initial magnetic field configurations (see \citealt{narayan2012} \& \citealt{sadowski2013}).  In the SANE case, the magnetic field is initialized with alternating poloidal loops, while the MAD initial field consists of a single poloidal loop, which results in the magnetic field playing a more dominant role in the dynamics of the disk.  We devise criteria to locate regions of field reversal and characterize the properties of the plasma in these regions.  We focus on the plasma-$\beta$ and magnetization parameter $\sigma$, which have been shown to play an important role in particle acceleration.  We also identify field components that are orthogonal to the reversing field, often referred to as guide fields, and quantify their strengths.  Our results will guide future particle-in-cell (PIC) studies of low-luminosity accretion flows.  Finally, we compute the time-dependent magnetic energy available in reconnection regions to assess whether this is a plausible mechanism to generate the observed X-ray variability of Sgr~A*.

\section{Characterizing Potential Reconnection Regions in MHD Simulations}

Magnetic reconnection takes place in regions where there is a reversal of 
magnetic field over a short characteristic length scale, in which the 
current density becomes large. In typical simulations of the local 
dynamics of reconnection, the initial condition is specified in terms of 
a Harris sheet, which has the magnetic field profile
\begin{equation}
\vec{B}=B_{0}\tanh{\frac{x}{L}}\hat{y}.
\end{equation}
In this geometry, the $y$-component of the magnetic field reverses
direction over a characteristic length $L$ in the $x$-direction.
This field reversal has a high curl associated with it, leading to a
sudden peak in the current density, which scales as
\begin{equation}
\vec{J}=\frac{B_{0}}{L}\sech^{2}{\frac{x}{L}}\hat{z}.
\end{equation}

There are only a small number of parameters that determine the particle heating and acceleration that results from reconnection events. These are the magnetization parameter
\begin{equation}
\sigma \equiv \frac{B^{2}}{4 \pi \rho c^2}, 
\end{equation}
which is the ratio of magnetic energy density to rest mass energy
density, and the plasma-$\beta$ parameter
\begin{equation}
\beta \equiv \frac{P_{\rm{gas}}}{P_{\rm{magnetic}}}= \frac{8 \pi n k T}{B^2}, 
\end{equation}
which specifies the ratio of gas pressure to magnetic pressure.  Here,
$\rho$, $n$, and $T$ are the mass density, number density, and temperature of the plasma
particles, respectively.

Another important quantity to consider for magnetic reconnection is
the magnitude and direction, if present, of the so-called guide field.
This is the component of the
magnetic field in the sheet perpendicular to the reconnecting field.
The effect of such a guide field on particle acceleration has been
studied in certain regimes (\citealt{wangh2016};
\citealt{dahlin2016}; \citealt{stanier2016}) and, in some cases, can have an effect on the resulting electron energy distribution.  

Our first goal is to devise an algorithm that will allow us to identify the
location and relevant properties of potential reconnection regions, i.e.,
Harris sheets, in global GRMHD simulations, which we describe in 
the following.

\section{Finding and characterizing current sheets}
  
As an illustrative example, we use two 60~hr (about $11,000 \; GM/c^{3}$) long GRMHD simulations of a radiatively
inefficient accretion flow onto a black hole (\citealt{narayan2012}; \citealt{sadowski2013}) that were performed using the HARM code \citep{gammie2003}.  These simulations were employed in a large study of the broadband, time-dependent emission from Sgr~A*  (\citealt{chan2015b}, b) where we coupled HARM to the radiative transfer algorithm GRay \citep{chan2013} and varied the black hole spin, density normalization, observer inclination, initial magnetic field configuration, as well as the electron thermodynamic prescription.  From these investigations, we identified 5 models that best fit the steady-state broadband spectrum as well as the previously observed 1.3 mm image size of Sgr~A* and also characterized their variability properties.
In the present study, we use two representative models from~\citet{chan2015a}: a SANE model with a black hole spin $a=0.7$ and a MAD model with a black hole spin $a=0.9$.  In general, the thermal SANE models tend to show short-lived, high amplitude variability in their IR and mm flux, while the thermal MAD models tend to show lightcurves dominated by smooth and long-timescale flux changes (\citealt{chan2015b}).

Our goal is to identify in each snapshot from these simulations potential regions of reconnection. Because
of the large shear in the accretion flow, the magnetic fields are
primarily toroidal and the alternating components occur primarily in
the azimuthal direction.  For this reason, we search through the
simulation volume for cells that have both a high current relative to
the mean value in the snapshot as well as very low values of the azimuthal component of the magnetic field $B_\phi$ in order to pick out the sheets where reconnection may occur.

Specifically, we consider 2D slices of the simulation volume at each
azimuthal angle $\phi$ at each snapshot and identify the points
that ($i$) have current magnitudes $\sqrt{J_\mu J^\mu}$ that are higher
than four times the mean current of that snapshot and ($ii$) $\phi$-components
of the magnetic field smaller than a fiducial value, characterized by the usual magnetization parameter $\sigma_{\phi}=B_{\phi}^{2}/(4 \pi m n c^{2})$.  We use a $\sigma_{\phi}$ threshold of $10^{-6}$.  We then apply an
algorithm similar to the one described in \citet{zhdankin2013} for identifying and
analyzing the statistics of current sheets in shearing box simulations
of MHD turbulence. For every point with grid indices ($i,j$) on an azimuthal slice of our domain picked out by the above criteria with current magnitude $J_{\rm{peak}}$, we consider all
4 adjacent points in the grid. If the current at an adjacent point is
above $J_{\rm{peak}}/2$, while also satisfying $\sigma_\phi<10^{-6}$, we consider it as
part of the same current sheet. We continue this process of scanning every
point in the sheet, considering all neighboring points, and applying
these criteria to them until no more points are being added to the
sheet.

\begin{figure}[htp]
\subfloat{
	\includegraphics[clip, width= \columnwidth]{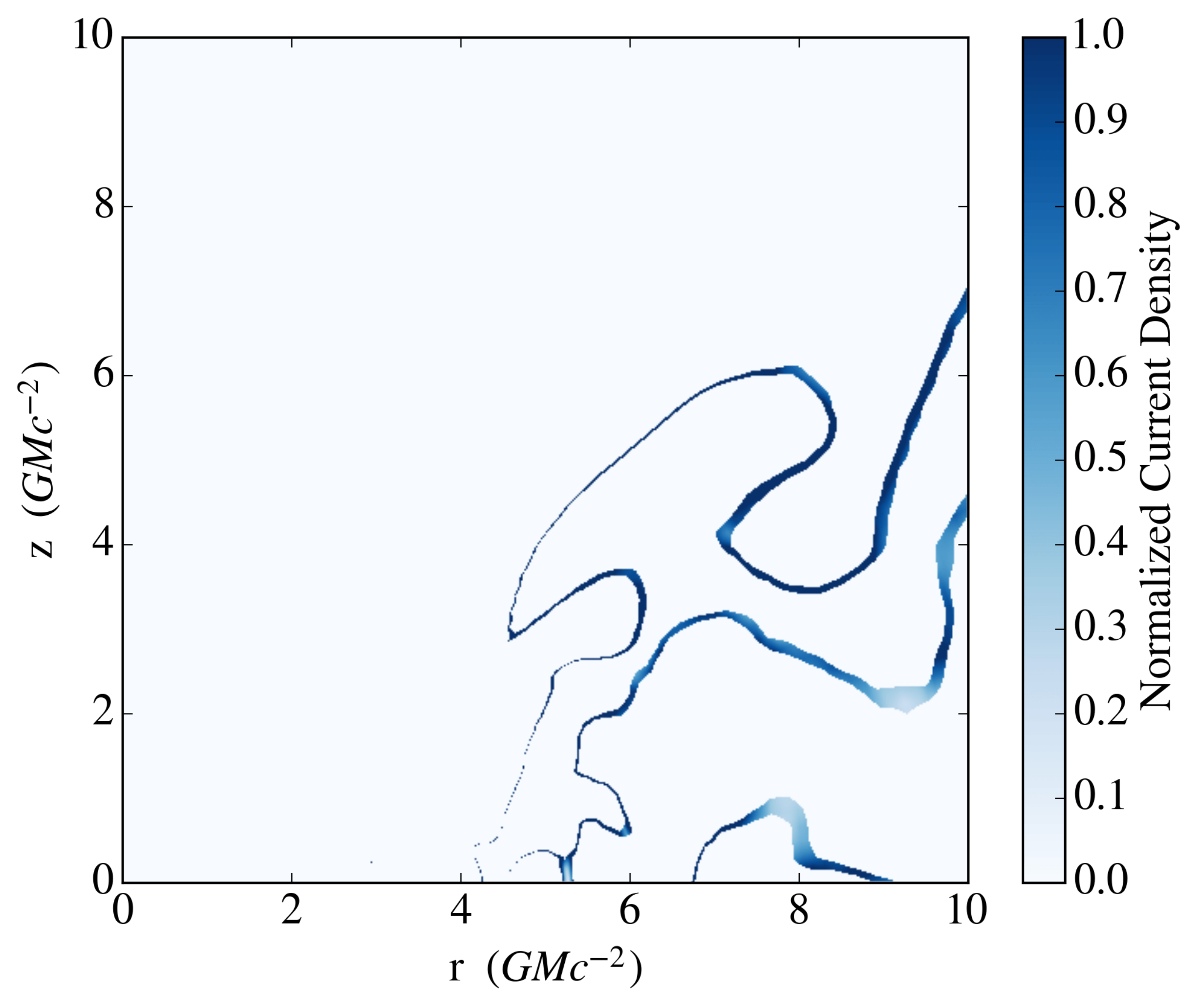}
}	

\subfloat{
	\includegraphics[clip, width= \columnwidth]{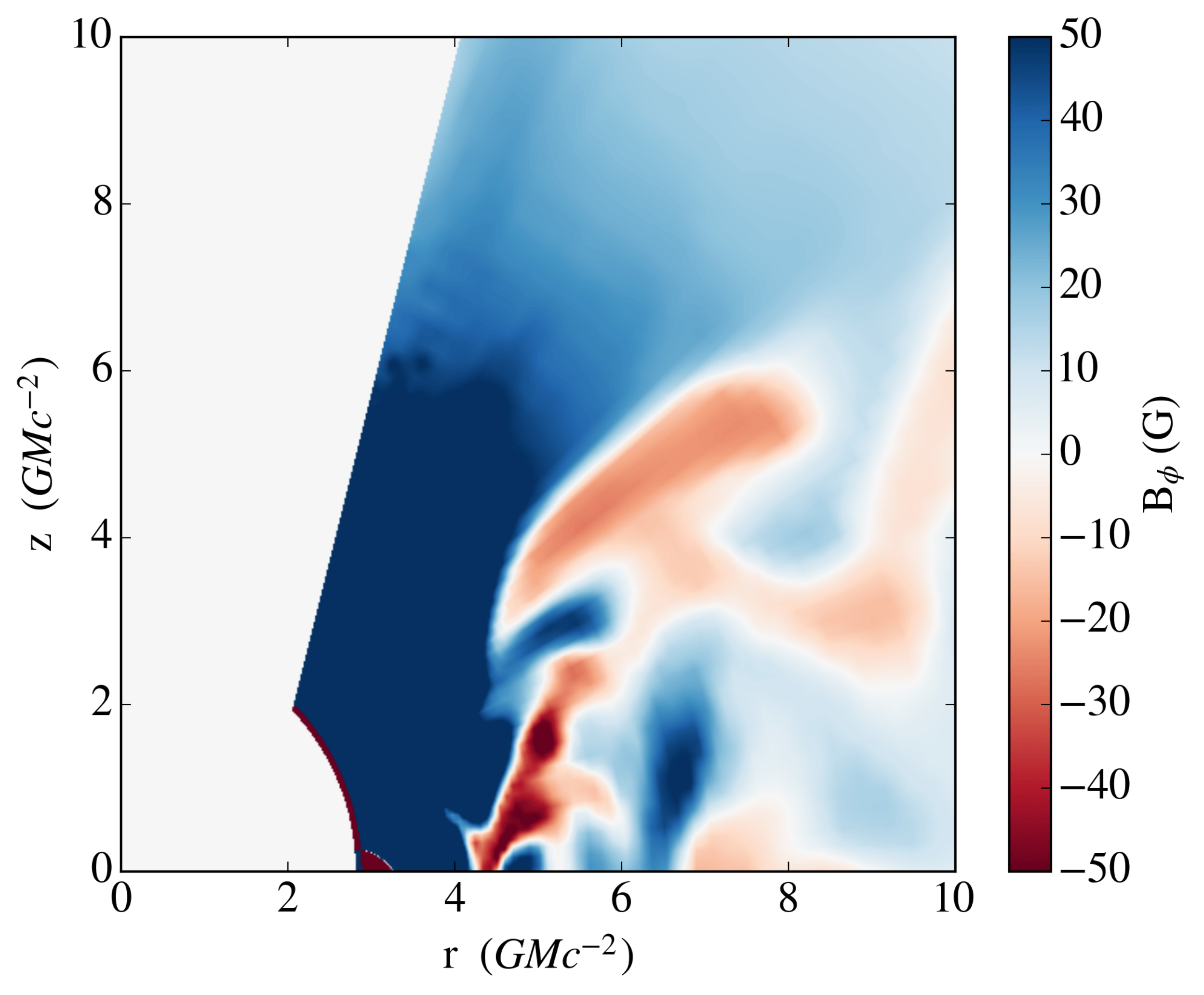}
	}

\caption{(Top) The current sheets picked out by our algorithm for the slice shown below, at the interface between regions of opposing magnetic flux.  Regions near the pole and within the ISCO are excised to avoid known numerical issues related to the density floor imposed.  (Bottom) A 2D slice of the azimuthal magnetic field in one snapshot of the SANE simulation, showing the presence of numerous opposing flux tubes that provide potential sites of reconnection}
\label{bphi_sheet_example}
\end{figure}

In the top panel of Figure~\ref{bphi_sheet_example}, we show the
result of applying this algorithm on a snapshot of the SANE simulation, with the $B_\phi$
configuration shown in the bottom panel of the same figure.  It is evident
that the regions between flux tubes of opposing azimuthal magnetic
flux are effectively picked out. When we repeat this procedure on all
adjacent azimuthal slices, we find that the current sheets show large
azimuthal extents throughout the flow, providing ample surface area
for neighboring flux tubes to reconnect over.

\subsection{Sampling The Plasma Properties Associated with Current Sheets}
Once we identify the current sheets in each snapshot, we characterize
the plasma parameters of these sheets that are relevant to magnetic
reconnection.  The location that we want to measure these parameters
at is not in the sheet itself but where the magnetic field reaches
its asymptotic value some distance away from the sheet in a direction
perpendicular to it.  This breaks down into two problems: finding the
direction perpendicular to the sheet and determining how far to go
along this direction until the magnetic field reaches its appropriate asymptotic
value.

In order to approximate the direction perpendicular to the current
sheet at a point ($i,j$) that has been flagged as belonging to the sheet, we
first find the local slope of the sheet about this point.  To
do this, we consider a box around each point ($i,j$) in the sheet, with width S+1, whose corners
are at ($i \pm S/2 , j \pm S/2$).
We use a value of $S=10$ pixels, which is generally smaller than the
radius of curvature of a current sheet.  We then calculate the slope
from the point in the center of this box ($i,j$) to every other point
($i^{\prime},j^{\prime}$) in the box which is flagged as being part of the current
sheet.  Taking the inverse tangent of this slope gives the angle
with respect to the horizontal of the line that passes through ($i,j$)
and ($i^{\prime}$,$j^{\prime}$).  We calculate the average of these angles, approximating the
angle of the current sheet about point ($i,j$) as

\begin{equation}
\theta_{mean}=\frac{1}{N}\sum_{\left(i^{\prime},j^{\prime}\right)_{n=1}}^{\left(i^{\prime},j^{\prime}\right)_{N}}
\arctan{\left[\frac{z\left(j^{\prime}_{n}\right)-z\left(j\right)}{r\left(i^{\prime}_{n}\right)-r\left(i\right)}\right]}
\end{equation}

We then calculate the mean slope:
\begin{equation}
m_{\rm{mean}}=\tan{\theta_{\rm{mean}}}
\end{equation}
and take the direction perpendicular to this slope as
\begin{equation}
m_{\perp}=-\frac{1}{m_{\rm{mean}}}
\end{equation}

We sample the plasma properties at some distance along
the normal where the toroidal magnetic field has reached its asymptotic value.
We approximate this location by scanning along the normal direction, given by equation~(7),
until the field profile flattens out.  We consider the field
sufficiently flat when the fractional change in magnetic field from
one computational cell along the normal to the next is less than three
percent, averaged over two adjacent cells\footnote[1]{Even though the approach outlined here for the definition of orthogonal directions is valid only for a flat spacetime, it is adequate for our present purposes both because we deal with short distances ($\sim0.1 M$) away from the current sheets and because we are interested in quantifying the typical values of the asymptotic magnetic field without being very sensitive to the precise direction.}.

\begin{figure}[t]
\includegraphics[width =0.5\textwidth]{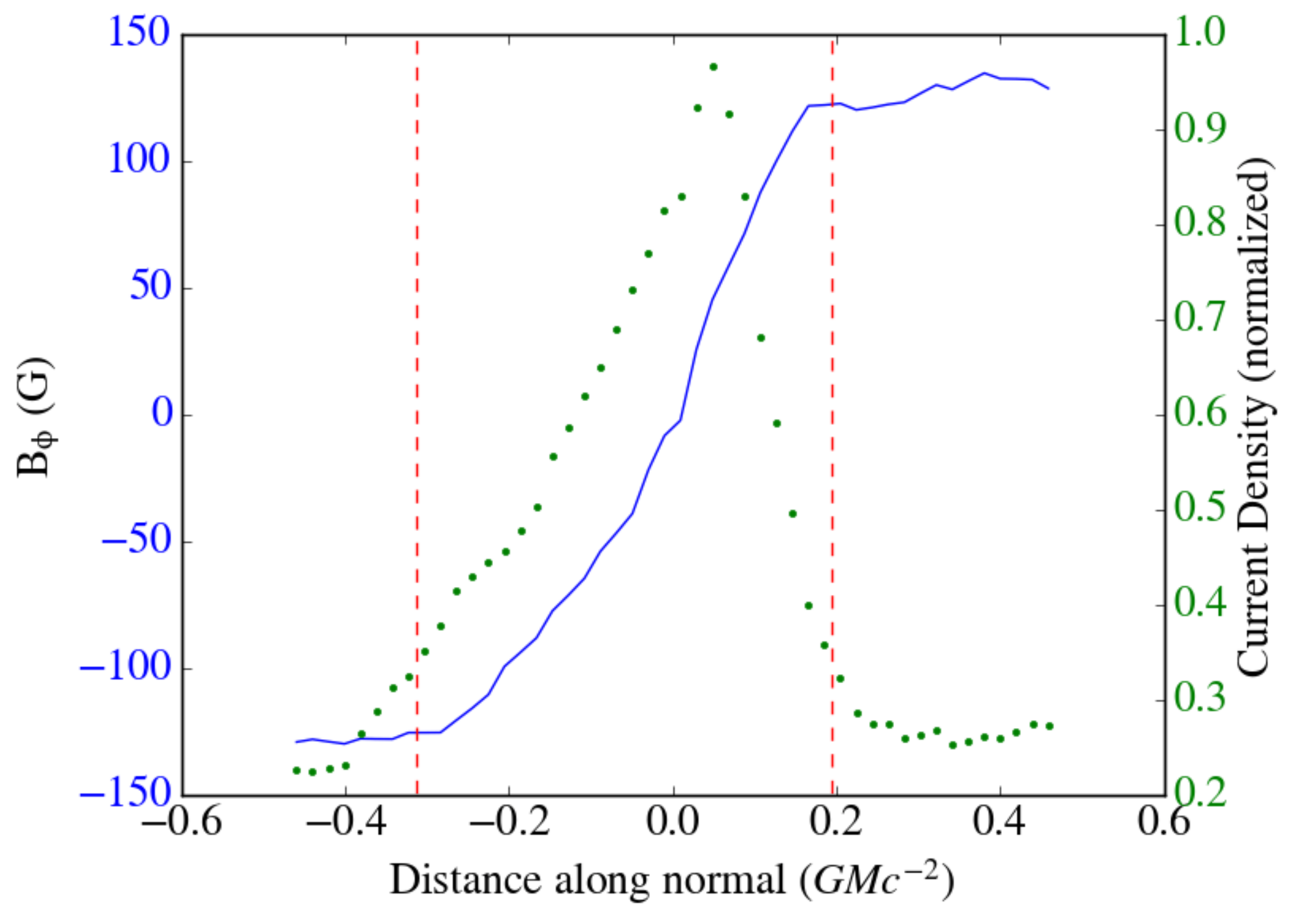}
\caption{Magnetic field (blue line) and current density (green points) profiles along the normal
  direction to a current sheet, showing the typical field reversal
  across the current sheet and the associated maximum in current
  density.  The red dashed lines indicate the location at which we
  sample the relevant plasma parameters.}
\label{field_profile}
\end{figure}

\begin{figure}[t]
\includegraphics[width=0.5\textwidth]{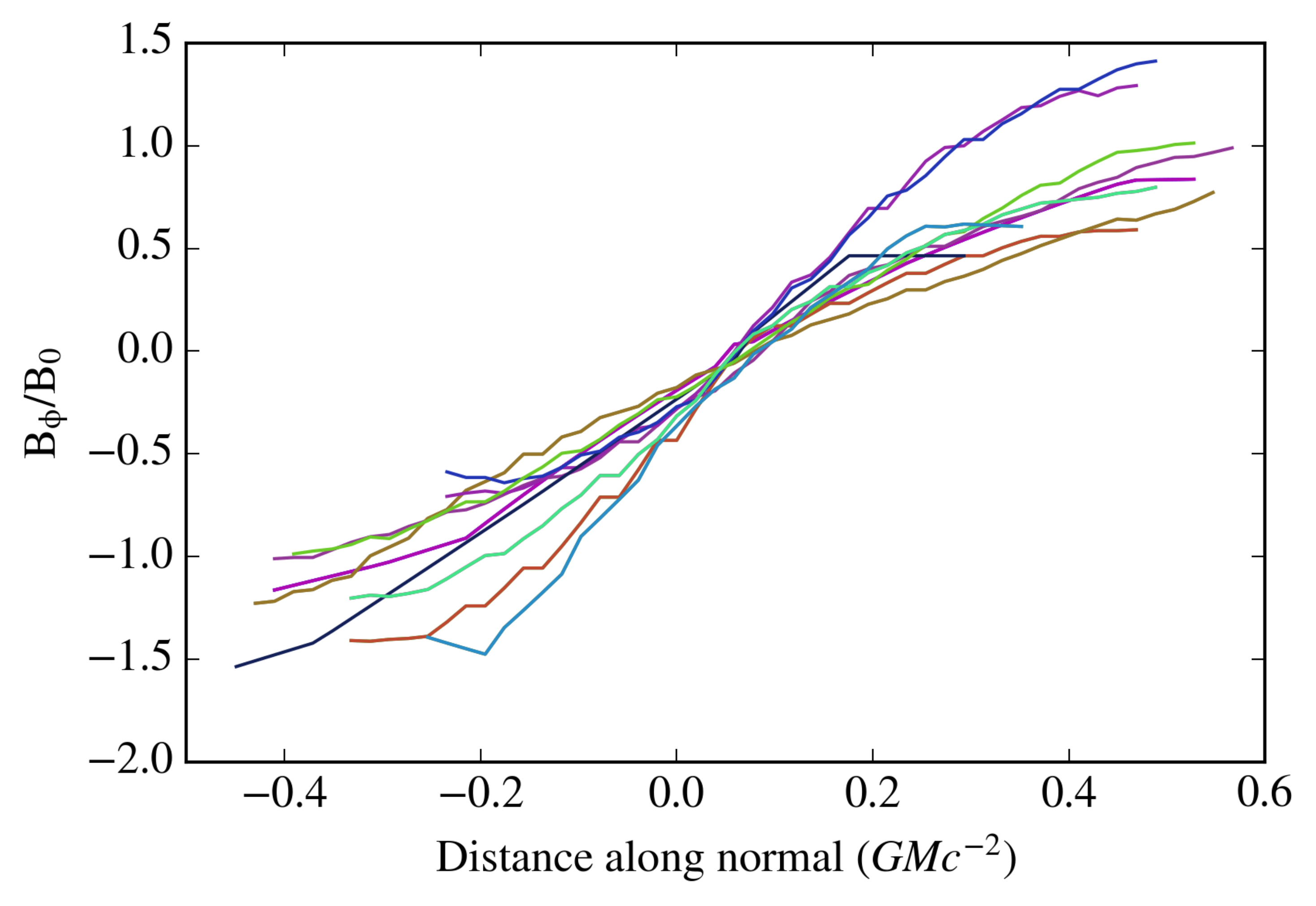}
\caption{A random selection of field profiles from the simulation across current sheets, showing that the typical behavior is reminiscent of the idealized Harris sheet structure, passing linearly through 0 and asymptoting to similar values on either side of the sheet.  The vertical scale is normalized to the average asymptotic magnetic field for each sheet.}
\label{rand_b}
\end{figure}

We show in Figure \ref{field_profile} the magnetic field and
current density profile along the normal of a current sheet
picked out by our algorithm.  We indeed see a Harris-sheet-like structure, with a magnetic field profile that passes through 0 and asymptotes to a fixed value at a distance $\approx$~0.2--0.3 $GMc^{-2}$ away from the center; the current density has the expected maximum associated with the steep gradient in magnetic field.  To illustrate the variety of current sheets and show their typical length scales and field profiles, we show in Figure \ref{rand_b} a randomly selected sample of current sheets identified in different snapshots and locations.  The magnetic field profiles again follow structures reminiscent of Harris sheets, with the magnetic field passing in a linear fashion through 0 and reaching an asymptotic value at a distance that is typically 0.2 to 0.6 $GMc^{-2}$ away from the center of the sheet.

\section{Plasma Properties of Current Sheets}

Having established the frequent occurrence and geometry of potential reconnection regions, our second goal is to investigate the properties of current sheets in time-dependent simulations of accretion flows and characterize the parameters relevant to non-thermal particle acceleration to inform further PIC studies.  We ultimately wish to determine the role of magnetic reconnection in contributing to the multiwavelength variability of low-luminosity accretion flows. 

Iterating through timesteps in our simulations, we find the
current sheets and, at every point in each sheet, determine the
asymptotic values of $\sigma$ and $\beta$ as well as the guide field strength at the center of the sheet for both our SANE and MAD simulations, as described below.  
\\
\subsection{Properties of SANE Current Sheets}
\begin{figure}[t]\includegraphics[width =0.5\textwidth]{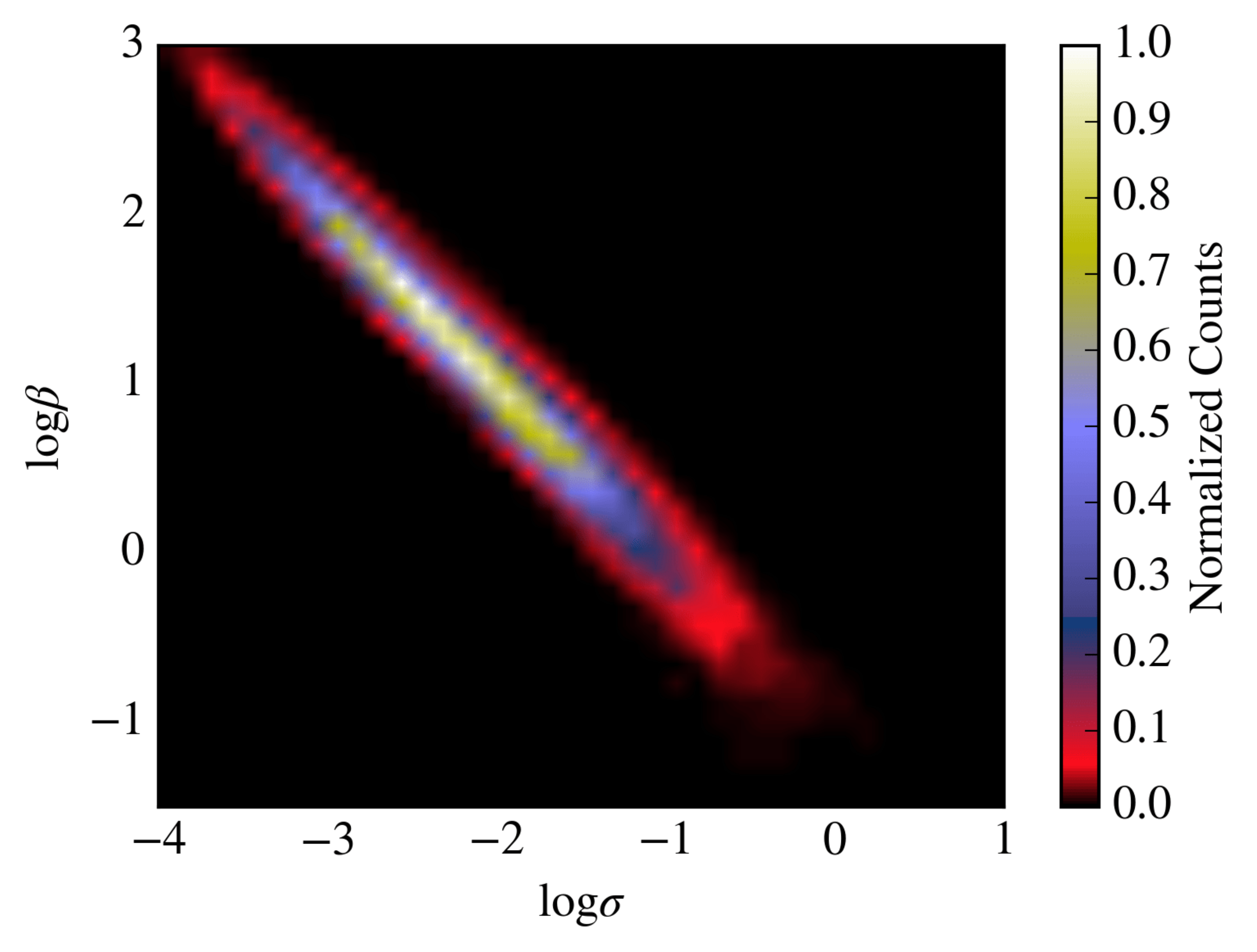}
	\caption{2d histogram of the magnetization $\sigma$ and plasma-$\beta$ across all current sheets 
		in the inner 10 $GMc^{-2}$ of the SANE simulation.}
	\label{SANE_2d_hist}
\end{figure}

For the SANE simulation, in the regions where reconnection may occur, the magnetization
$\sigma$ ranges from $10^{-4}$ to 1, while the plasma-$\beta$ ranges from 0.1 to $10^{3}$, as shown in Figure \ref{SANE_2d_hist}.    
\\
The anticorrelation evident
in Figure \ref{SANE_2d_hist} (see also Fig.~\ref{MAD_2d_hist}) occurs because the magnetization parameter $\sigma$ scales as $B^{2}/n$,
while the plasma-$\beta$ scales as $\left(B^{2}/n\right)^{-1}$.  The spread arises because the plasma-$\beta$ also depends on the plasma temperature. 

The most promising subspace of this region for particle acceleration to be efficient is the high-$\sigma$, low-$\beta$ (bottom-right) regime, where there is maximal magnetic energy to dissipate into the particles and fairly little gas pressure relative to the magnetic pressure, such that the plasma is magnetically dominated.  The inferred ranges of $\sigma$ and $\beta$ are interesting for a
number of reasons.  Studies have only recently begun for this transrelativistic regime \citep{werner2016} and the physics of particle acceleration and heating in these conditions are not yet fully understood.  While the ions in this regime remain
non-relativistic (because $\sigma$ is of order ~1), the electrons will
likely be accelerated (or heated on average) to highly relativistic speeds, since
$\sigma_{e} \equiv B^{2}/\left(4 \pi \rho_{e}c^{2}\right) = \sigma m_{i}/m_{e} \approx10^{3}$, which is an estimate of the characteristic electron Lorentz factors expected from reconnection.

\begin{figure}[!h]
	\includegraphics[width=.5\textwidth]{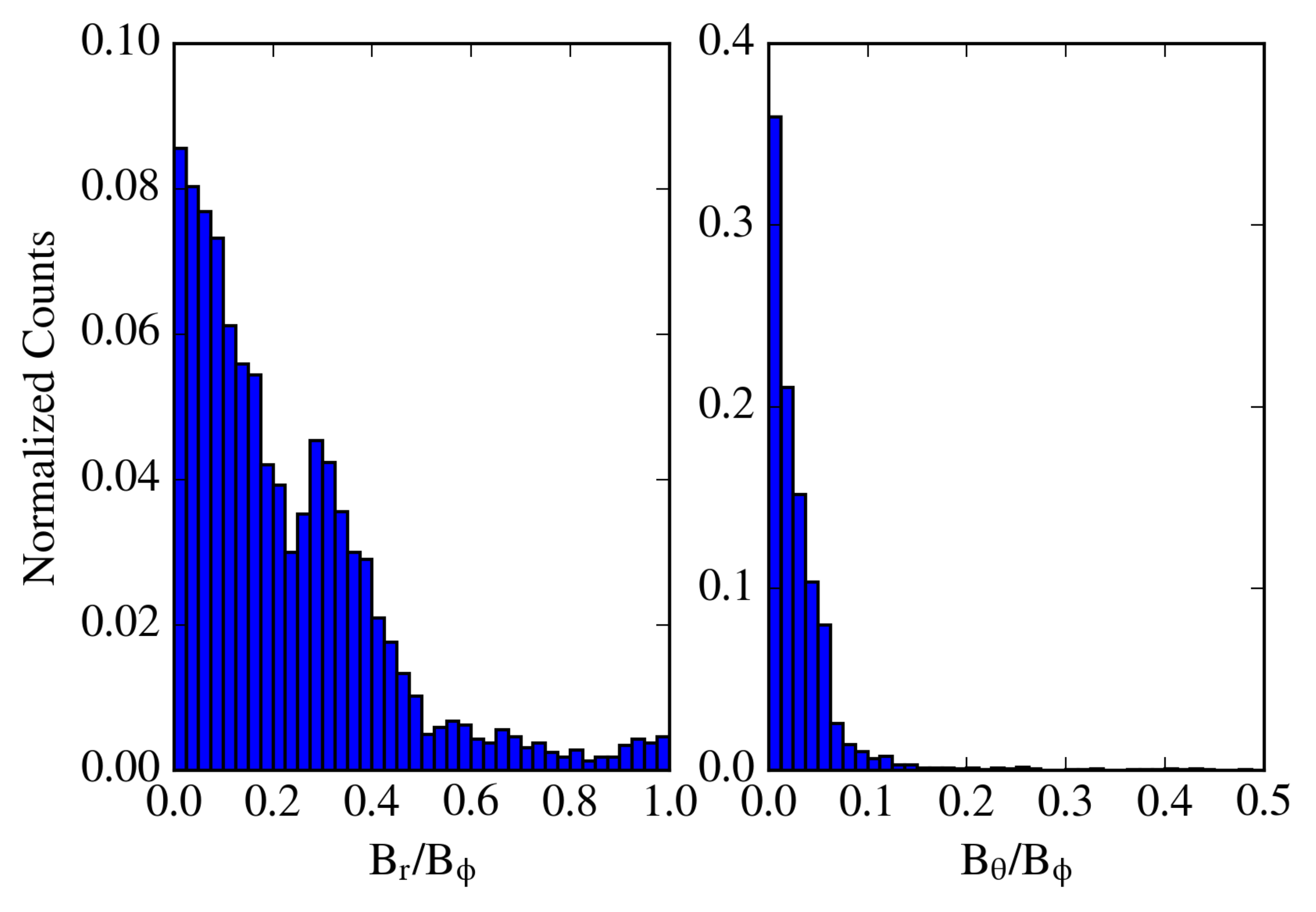}
	\caption{Histogram of guide fields in the SANE simulation, scaled to $B_{\phi}$, the component showing field reversal.  A large number of current sheets have no guide fields associated with them and, when present, the guide fields tend to be quite weak}
	\label{guide_field}
\end{figure}

Finally, in Figure \ref{guide_field} we show a histogram of the relative guide field strengths in the SANE simulation.  It is evident that both cases of weak ($B_{r}/B_{\phi} < 0.5$)  and of no guide fields are of interest for the purposes of these simulations.  Even weak guide fields may play an important and potentially adverse role in determining the outcome of magnetic reconnection and must be explored via PIC simulations in the transrelativistic regime.


\subsection{Properties of MAD Current Sheets}
\begin{figure}[t]
\includegraphics[width =0.5\textwidth]{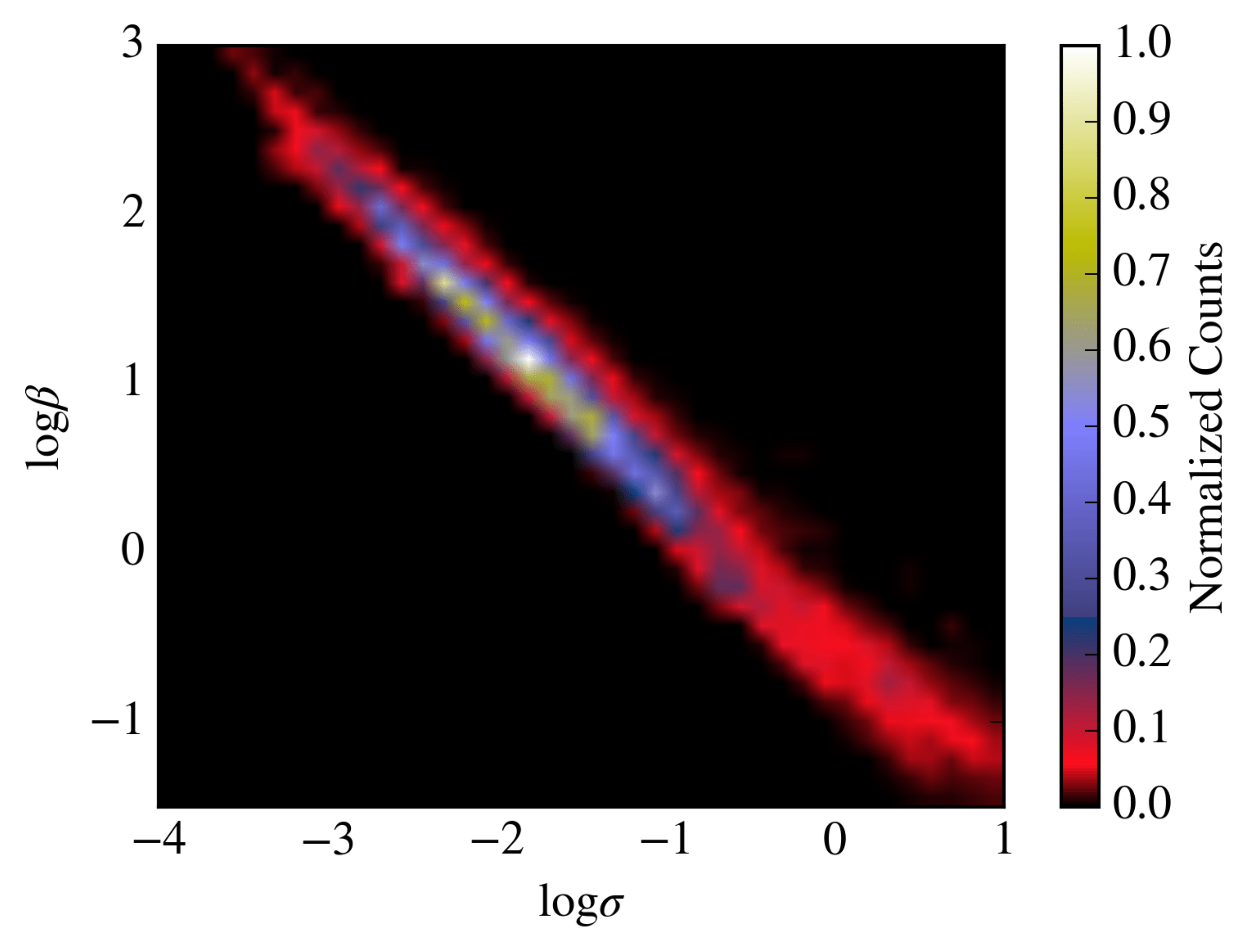}
\caption{2d histogram of the magnetization $\sigma$ and plasma$-\beta$ across all current sheets 
in the inner 10 $GMc^{-2}$ of the MAD simulation.}
\label{MAD_2d_hist}
\end{figure}

For the MAD simulation, we find that, in the regions of potential reconnection, $\sigma$ ranges from 10$^{-3}$ to 10, while $\beta$ ranges from 0.03 to 10$^{3}$, as shown in Figure \ref{MAD_2d_hist}.  This is roughly an order of magnitude higher (lower) than the $\sigma$ ($\beta$) values in the SANE simulation, hinting that particle acceleration may be more efficient in these systems.  

\begin{figure}[b]
\includegraphics[width=.5\textwidth]{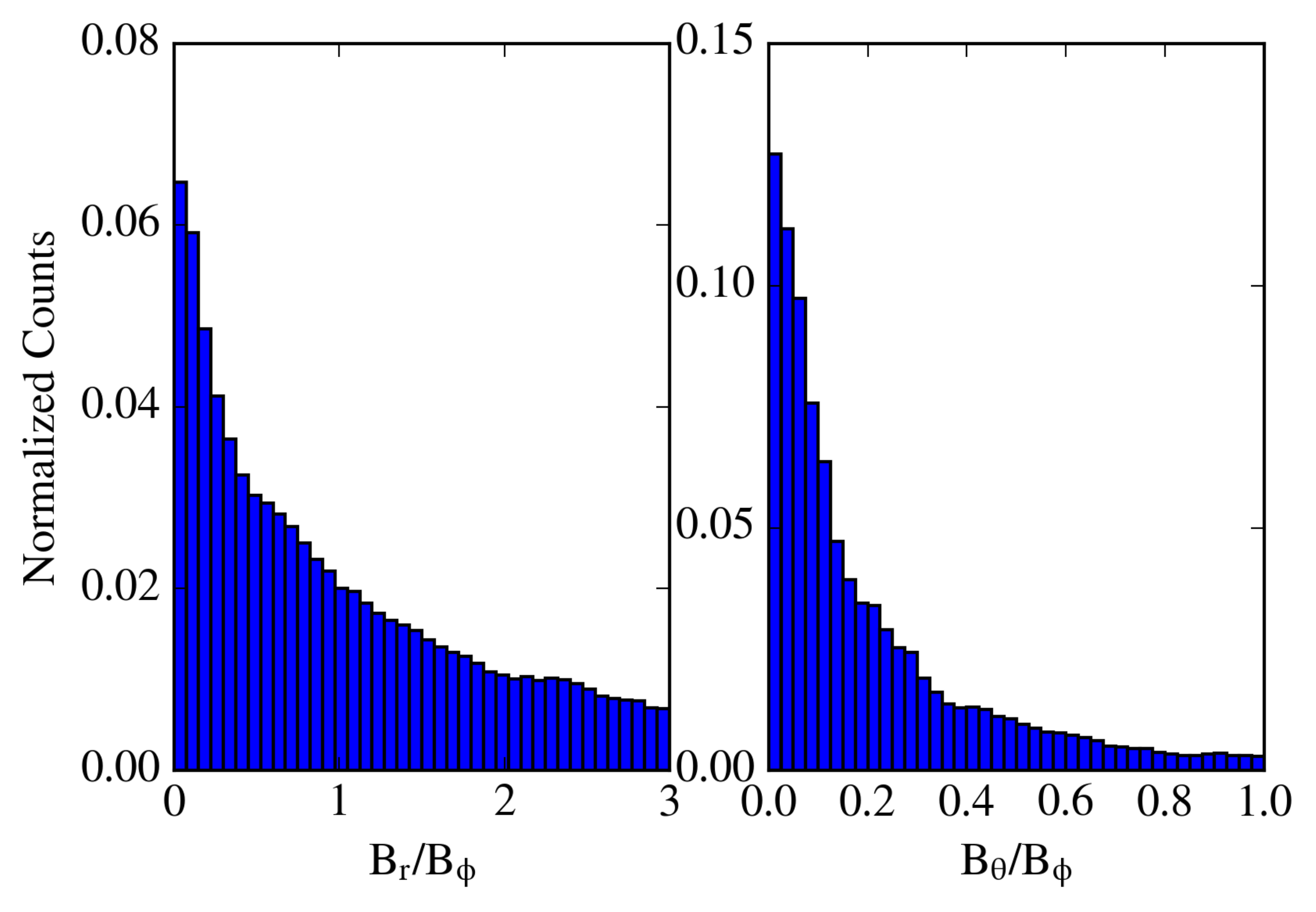}
\caption{Histogram of guide fields in the MAD simulation, scaled to $B_{\phi}$, the component showing field reversal.  While many sheets have little to no guide fields present, there are a significant number of current sheets with strong guide fields that will likely impact the efficiency of particle acceleration in these sheets.}
\label{MAD_guide_field}
\end{figure}

We show the guide fields in the MAD simulation in Figure \ref{MAD_guide_field}.  In stark contrast to the SANE guide fields, which are weak relative to the reconnecting field, the MAD guide fields are stronger and can be comparable to the reconnecting ones.  
While the typical values of the magnetization $\sigma$ and the plasma-$\beta$ are more favorable in terms of particle acceleration in the MAD simulations, the stronger guide fields may alter the outcome of the reconnection event for the particle distribution.

\section{Variability of Magnetic Energy Available For Reconnection}
\begin{figure*}
	\includegraphics[width=.5\textwidth]{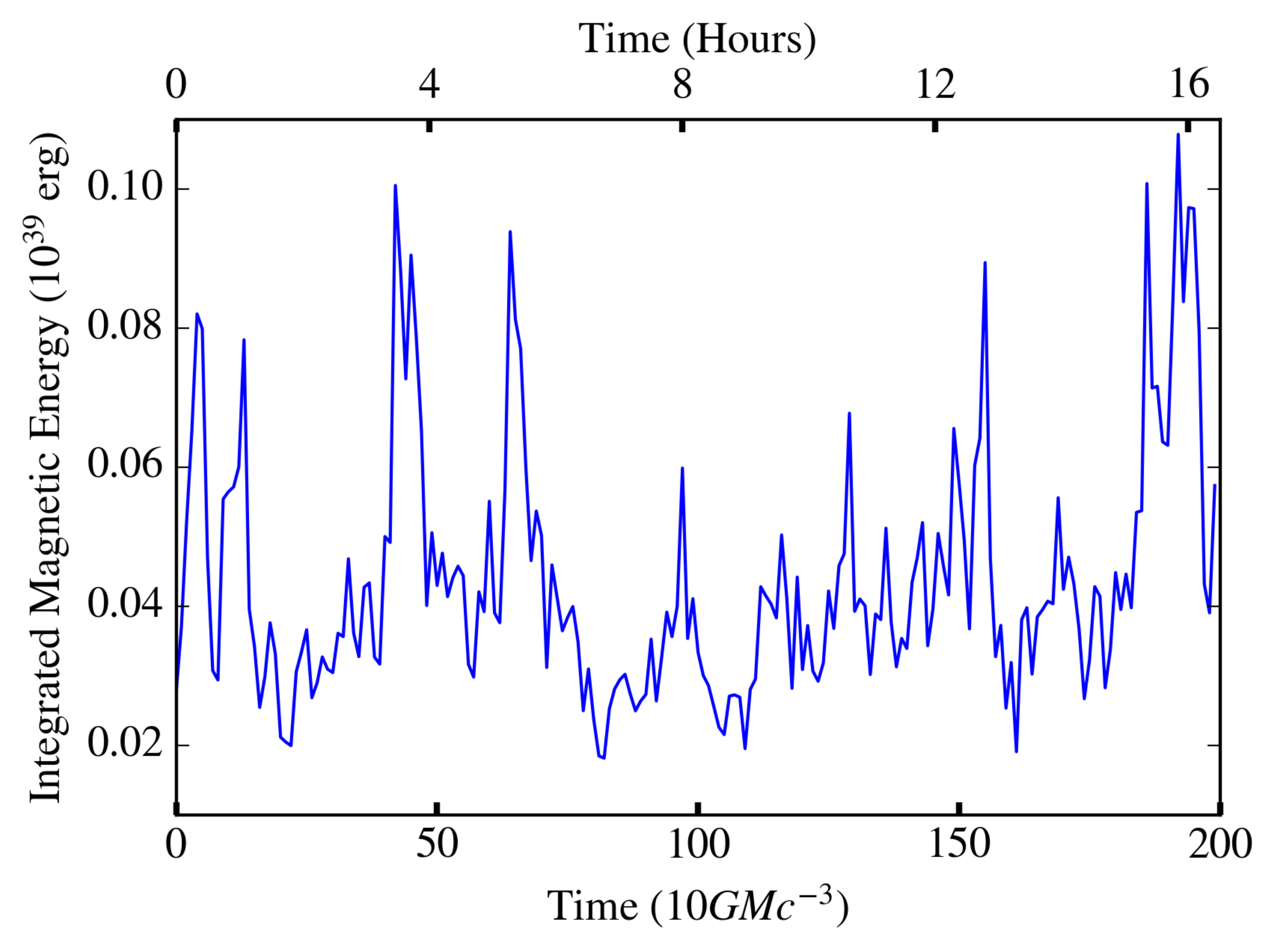}
	\includegraphics[width=.5\textwidth]{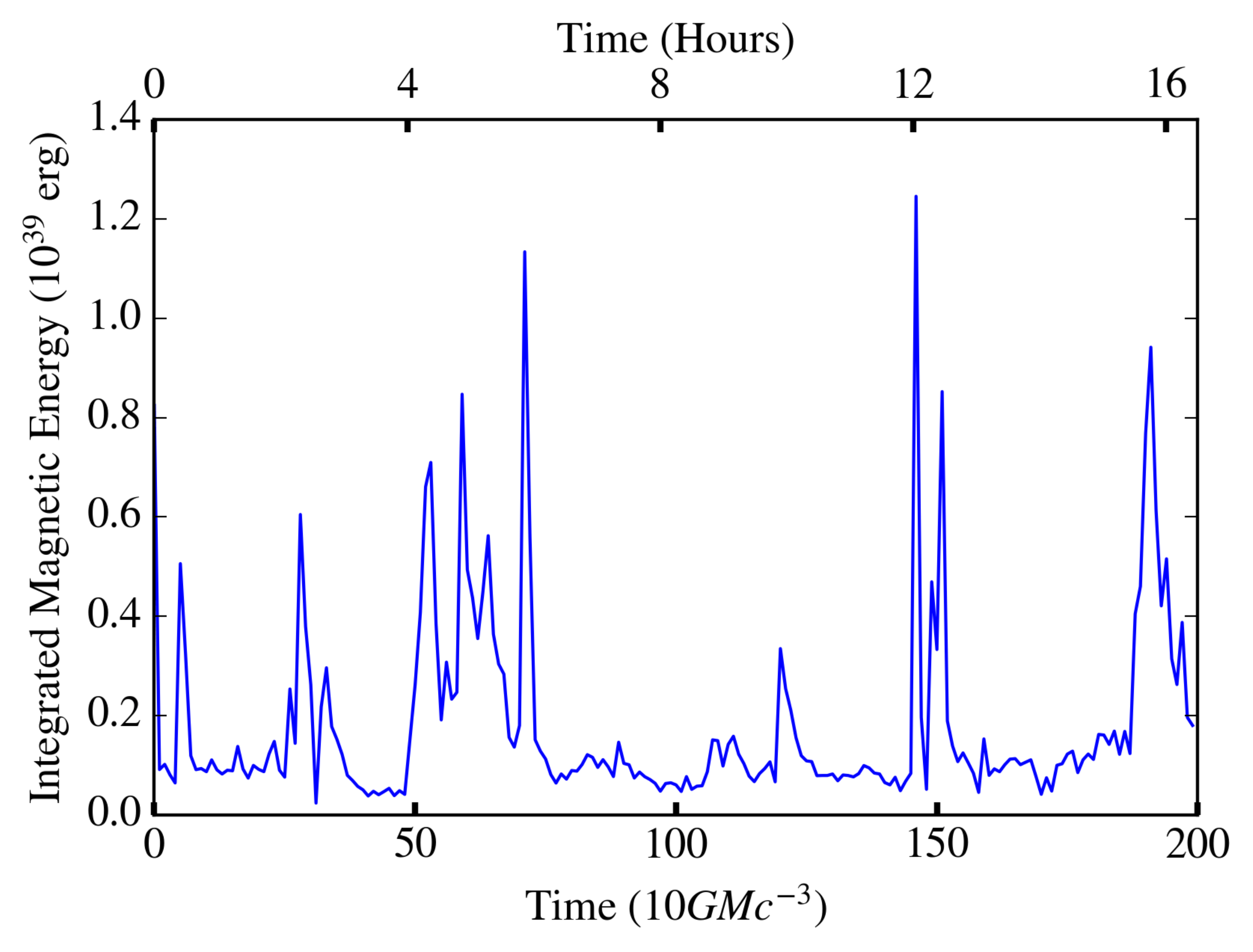}\label{MAD mag energy}
	\caption{Magnetic energy of reconnection regions as a function of time in the (\emph{left}) SANE and (\emph{right}) MAD simulations.  Note the rapid and strong variation over short timescales in both cases, making magnetic reconnection a promising candidate for contributing to the X-ray variability.}\label{mag energy}
\end{figure*}

We finally examine the time variability of energy available to reconnection
throughout the accretion flow.  One motivation for this is to assess whether reconnection events can contribute substantially to the high energy variability of low luminosity accretion flows, as has been extensively observed in the case of Sgr~A*.  We integrate the magnetic energy density,
$B^{2}/8\pi$, over the reconnecting volume bounded by the surfaces
defined by the asymptotic magnetic field location and obtain in this way the total
magnetic energy in the reconnection regions throughout the flow.  We
plot the results of this in Figure \ref{mag energy} for both the SANE and MAD simulations.

We see that the turbulent nature of the accretion flow very often leads to the
formation of transient current sheets that result in a highly
time-varying magnetic energy being available to reconnection.  This indicates that magnetic reconnection likely is a significant
contributor to the variability of such systems.  The SANE model produces persistent variability due to the high levels of turbulence in the disk.  The MAD system has less turbulence and, hence, fewer variations, but the higher degree of magnetization means that, when a current sheet does develop, it typically has more magnetic energy associated with it.  For this reason, we find that the MAD simulation is characterized by fewer but stronger variations in the magnetic energy available to reconnection.  

Including the variability properties of non-thermal electrons that are accelerated in these current sheets will likely alter significantly the earlier finding of \citet{chan2015b}, who used models that assumed a purely thermal electron distribution.  In that early work, the variability of MAD simulations was characterized by smooth long timescale variations in the flux.  It is clear, however, from the present  analysis of the MAD simulation that there is the potential for having a sudden injection of non-thermal electrons associated with the spikes in Figure \ref{MAD mag energy}, which can then result in corresponding flares in the lightcurve.

Based on previous studies, we expect some fraction of this magnetic energy to
go towards accelerating particles.  This acceleration efficiency will, in principle,
depend on the flow conditions, such as $\sigma$ and $\beta$, and can
be found through PIC simulations.  In order to go from a picture such
as the one shown in Figure~\ref{mag energy} to the non-thermal particle energy
as a function of time, the magnetic energy as a function of time must
be combined with the acceleration efficiency as a function of flow
parameters, which will likely result in even more dramatic variation
of energy on short timescales.  

To estimate whether the energy available for reconnection is a plausible explanation for flares of these magnitudes, we calculate the total energy from an average X-ray flare from Sgr~A*.  Observations of X-ray flares from Sgr~A* show typical luminosities from $\sim 10^{34}$ erg~s$^{-1}$ to $2 \times 10^{35}$ erg~s$^{-1}$ and typical timescales from hundreds of s to 8 ks \citep{neilsen2013}.  With a luminosity of $5 \times 10^{34}$ erg~s$^{-1}$ and a duration of 1000 s, about $5 \times 10^{37}$ erg is being released in a typical flare.  

Considering Figure \ref{mag energy}, we see for the SANE model that the energy available to reconnection peaks at typical values around $10^{38}$ erg, while typical MAD energies are an order of magnitude higher than this.  This shows that there is enough energy available to reconnection in these simulations to plausibly account for the observed energy released during these flares.  Moreover, the efficiency $\xi$ that determines the fraction of magnetic energy that goes into particle acceleration must be quite high in the case of SANE models, which have typically lower magnetizations and hence less magnetic energy associated with their current sheets.

In \citet{ball2016}, we characterized the non-thermal particle distribution using $\eta$, the fraction of non-thermal to thermal energy densities in the fluid and power-law index, p.  We found that significant X-ray flares can occur while satisfying the observed quiescent X-ray constraints for values of $\eta = 0.1$ and a conservative power-law index of $p=-3.5$.  To connect our present results to these earlier findings, we express $\eta$ in terms of $\beta$ and $\xi$ as 
\begin{equation}
\eta \equiv \frac{E_{nt}}{E_{th}} = \frac{\xi B^{2}}{nkT8\pi} = \xi \beta^{-1}.
\label{beta_constraint}
\end{equation}
We can rewrite this as a constraint on the plasma-$\beta$ using the $\eta$ found in our previous study to result in significant X-ray variability, i.e.,
\begin{equation}
0.1 \left(\frac{\eta}{0.1}\right) \beta \leq \xi.
\end{equation}
This places a constraint on the plasma-$\beta$, given a local $\xi$, which must be found as a function of flow parameters via PIC simulations.  Note that $\xi$, by definition, cannot be greater than 1, placing a strict upper limit of $\beta=10$ for regions where there is sufficient magnetic energy to accelerate particle to the energies required to generate the flux excursions demonstrated in \citet{ball2016}.  More realistically, $\xi$ is likely to be of order 0.1, resulting in an upper limit of $\beta\approx1$.  As shown in Figure \ref{SANE_2d_hist}, we find that we indeed identify many current sheets satisfying this condition.

\section{Conclusions}
In this paper, we investigated the detailed structure of current sheets and their
plasma properties in GRMHD simulations of radiatively inefficient accretion flows.  We found that the
regimes of plasma parameters relevant to magnetic reconnection have
been relatively unexplored in terms of non-thermal particle
acceleration.  Specifically, we found that the magnetization $\sigma$ in the vicinity of current sheets in the SANE simulation is of order
$10^{-4}$ to 1 , while the plasma-$\beta$ is of order 0.1 to
$10^{3}$.  Current sheets in the MAD simulation have magnetization $\sigma$ ranging from $10^{-3}$ to 10 and plasma-$\beta$ from 0.03 to $10^{3}$.  Additionally we find that, in these regions, there is a relatively small spread in temperature, leading to a tight correlation between the parameters $\sigma$ and $\beta$.  We also characterized the guide fields found in current sheets,
which can play a role in governing the details of particle
acceleration, and found that the ratio of guide field to reconnecting
field strength is typically 0--0.5 for SANE simulations, but can be of order unity  in MAD simulations.  

GRMHD simulations need to use subgrid models in order to
account for physical effects that cannot be resolved or incorporated in MHD.  In order to employ correctly subgrid models of reconnection, we must improve our
understanding of particle acceleration and heating in the parameter
space we lay out here.

In addition to characterizing the plasma properties of current sheets,
we also calculated the magnetic energy available to reconnection throughout the simulations.  We found that the turbulent
nature of the accretion flow leads to current sheets of varying
characteristics continuously forming and dissipating in the flow.  This leads to
a highly variable amount of energy available to reconnect and
dissipate into heating and particle acceleration and makes magnetic
reconnection a promising candidate for contributing to the X-ray
variability of Sgr~A* and other black holes with similar accretion characteristics.  Additionally, we found that there is indeed enough energy available to reconnection around current sheets to account for typical flares observed from Sgr~A*.  We conclude that if this mechanism is responsible for the X-ray flares, then the acceleration efficiency must be reasonably high for SANE disks and can be lower for the MAD model.

\acknowledgements
We gratefully
acknowledge support for this work from Chandra
Award No. TM6-17006X and from NASA TCAN
award NNX14AB48G. 
DP acknowledges support from the Radcliffe Institute for Advanced Study at Harvard University.
FO gratefully acknowledges a fellowship from the John Simon Guggenheim Memorial Foundation 
in support of this work.

\appendix 
\section{Calculation of Current Density in a Kerr Metric}
In order to find regions in our GRMHD simulation where magnetic
reconnection would occur, if it was explicitly included, we need to
identify regions of high current density.  However, GRMHD simulations
typically evolve only 4 quantities, i.e., the magnetic field, density, fluid
velocity, and internal energy, and the current density is typically not explicitly computed in
the simulation.  To understand the structure of
the current density throughout the flow, we
calculate it from the electromagnetic tensor,
\begin{equation}\label{maxwell tensor}
F^{\mu \nu}= \left( \begin{array}{cccc}
0 & -E_{1} & -E_{2} & -E_{3} \\
E_{1} & 0 & B_{3} & -B_{2} \\
E_{2} & -B_{3} &  0 & B_{1}\\ 
E_{3} & B_{2} & -B_{1} & 0 \end{array} \right)
\end{equation}

via

\begin{equation}\label{maxwell}
J^{\nu}=\nabla_{\mu}F^{\nu \mu}
\end{equation}
where $\nabla_\mu$ represents the covariant derivative.  

Breaking up the four-current into its zeroth and i$^{th}$ components
(henceforth, Greek indices run from 0-3, while Latin indices go from
1-3), we can rewrite equation (\ref{maxwell}) as

\begin{equation}\label{max1}
\nabla_{j}F^{ij} - \nabla_{0}F^{0i}=J^{i}
\end{equation}
\begin{equation}\label{max2}
\nabla_{i}F^{0i}=J^{0},
\end{equation}
In the comoving frame, $F^{0 i} = 0$, equation (\ref{max2}) reads
$J^{0} = 0$. Additionally, the displacement current is 0
($\nabla_{0}F^{0i}=0$).  We can then write the three-current as
\begin{equation}\label{ji}
J^{i} = \partial_{j}F^{ij} + \Gamma^{i}_{j \lambda}F^{\lambda j} + \Gamma^{j}_{j \lambda}F^{i \lambda}.
\end{equation}
Considering the 2nd term on the right hand side, $\Gamma^{i}_{j
  \lambda}F^{\lambda j}$, we see that for every term in the sum, the
symmetry of $\Gamma^{i}_{jk}$ about its lower two indices and the anti-symmetry of
$F^{\mu \nu}$ implies perfect cancellation.  Hence, equation (\ref{ji})
reduces to

\begin{equation}\label{ji2}
J^{i} = \partial_{j}F^{ij} + \Gamma^{j}_{j \lambda}F^{i \lambda}.
\end{equation}

We use the Christoffel symbols for the metric of an
uncharged Kerr black hole, which in Boyer-Lindquist coordinates is given by
\begin{equation}\label{BL metric}
\begin{aligned}
ds^{2} = -\frac{\Delta}{\Sigma}\left(dt - a \sin^{2}\theta d\phi \right)^{2} + \frac{\Sigma}{\Delta}dr^{2} 
+ \Sigma d\theta^{2} \\+ \frac{\sin^{2}\theta}{\Sigma}\left(\left(r^{2} + a^{2}\right)d\phi + a dt \right)^{2} 
\end{aligned}
\end{equation}
with 
\begin{equation}
\begin{aligned}
\Delta=r^{2} - 2Mr + a^{2} \\
\Sigma = r^{2} + a^{2}\cos^{2}\theta. 
\end{aligned}
\end{equation}
\\ Here, M and a are the mass and angular momentum per unit mass of the black hole.

For this metric, the relevant nonzero
Christoffel terms in equation (\ref{ji2}) are $\Gamma^{1}_{11} , \Gamma^{1}_{1 2} ,
 \Gamma^{3}_{3 1} , \Gamma^{3}_{3 2}   
 ,  \Gamma^{2}_{2 1}  ,\Gamma^{2}_{2 2}.$
With these, we can write out the individual components of $J^{i}$ as

\begin{equation}
J^{1} =  \partial_{j}F^{1j} + \left(\Gamma^{1}_{12} + \Gamma^{3}_{32} + \Gamma^{2}_{22}\right)F^{12} ,
\end{equation}

\begin{equation}
J^{2} = \partial_{j}F^{2j} + \left(\Gamma^{1}_{11} + \Gamma^{3}_{31} + \Gamma^{2}_{21}\right)F^{21} ,
\end{equation}
and
\begin{equation}
\begin{aligned}
J^{3} =  \partial_{j}F^{3j} + \left(\Gamma^{1}_{11} + \Gamma^{2}_{21} + \Gamma^{3}_{31}\right)F^{31} \\+ \left(\Gamma^{1}_{12} 
+ \Gamma^{3}_{32} + \Gamma^{2}_{22}\right)F^{32}.
\end{aligned}
\end{equation}
Using these equations, we solve for the current density in GRMHD
simulations at every timestep and identify current sheets where reconnection may take
place.

\bibliographystyle{apj}
\bibliography{david_bib}

\end{document}